\documentclass[
 reprint,
superscriptaddress,
 amsmath,amssymb,
aps,
prd,
twocolumn
]{revtex4-2}

\usepackage{amsfonts,amssymb,amsmath}
\usepackage[caption=false]{subfig} 
\usepackage{graphicx, hyperref, tabularx, dcolumn}
\usepackage{mathtools}

\usepackage{color}
\usepackage[dvipsnames]{xcolor}
\usepackage{soul}
\usepackage{tensor, comment}
\usepackage{enumitem}
\usepackage[export]{adjustbox}
\usepackage{cleveref}
\usepackage{orcidlink}

\begin{document}

\preprint{APS/123-QED}

\title{Unveiling the electrodynamic nature of spacetime collisions}

\author{Siddharth Boyeneni}\thanks{These authors contributed equally to this work.}
\affiliation{TAPIR, Mailcode 350-17, California Institute of Technology, Pasadena, CA 91125, USA}
\author{Jiaxi Wu\,\orcidlink{0000-0003-3829-967X}}\thanks{These authors contributed equally to this work.}
\affiliation{TAPIR, Mailcode 350-17, California Institute of Technology, Pasadena, CA 91125, USA}
\author{Elias R. Most\,\orcidlink{0000-0002-0491-1210}}
\affiliation{TAPIR, Mailcode 350-17, California Institute of Technology, Pasadena, CA 91125, USA}
\affiliation{Walter Burke Institute for Theoretical Physics, California Institute of Technology, Pasadena, CA 91125, USA}

\begin{abstract}
Gravitational waves from merging binary black holes present exciting opportunities for understanding fundamental aspects of gravity, including nonlinearities in the strong-field regime. One challenge in studying and interpreting the dynamics of binary black hole collisions is the intrinsically geometrical nature of spacetime, which in many ways is unlike that of other classical field theories. By exactly recasting Einstein's equations into a set of coupled nonlinear Maxwell equations closely resembling classical electrodynamics, we visualize the intricate dynamics of gravitational electric and magnetic fields during the inspiral, merger, and ring-down of a binary black hole collision.
\end{abstract}


\maketitle



{\bf Introduction.} Binary black hole mergers are at the heart of modern gravitational wave science \cite{Barack:2018yly}. Since the initial detection of merging stellar-mass black hole binaries by the LIGO collaboration \cite{LIGOScientific:2016aoc}, more than one hundred binary black hole collisions have been detected \cite{LIGOScientific:2018mvr,LIGOScientific:2020ibl,KAGRA:2021vkt}.
The physics of binary black hole mergers has been well investigated using a variety of tools and approximations for the different stages of the merger: post-Newtonian theory \cite{Blanchet:2013haa} and effective one-body approaches \cite{Buonanno:1998gg} for the inspiral, and numerical relativity simulations for the merger \cite{Centrella:2010mx,Lehner:2014asa}. Recent insights into the nonlinearity of black hole ring-down \cite{Giesler:2019uxc,Cheung:2022rbm,Mitman:2022qdl} have reinvigorated theoretical explorations of black hole dynamics in general relativity.

Several attempts have been made to understand the highly
nonlinear nature of the merger
\cite{Lovelace:2014twa,Pook-Kolb:2019iao,Pook-Kolb:2020jlr,Pook-Kolb:2020zhm}.
One notable example is the use of gravitometrodynamics
\cite{Maartens:1997fg}, trying to describe the tendicity (tidal forces) and
vorticity (space-time drag) of a dynamical space-time \cite{Owen_2011}.
This approach has led to new insights into both inspiral
\cite{Keppel:2009tc} and ring-down of a binary black hole space-time
\cite{Nichols:2012jn}. On a mathematical level, gravitometrodynamics relies
on decomposing the curvature tensor into analogue electric and magnetic
fields, which obey Maxwell-like equations \cite{Maartens:1997fg}, including
the conservation of a super-energy \cite{bel1959introduction}. However,
these fields are not classical electric and magnetic fields as they, e.g.,
in the case of the electric field, represent tidal and not Coulomb forces.

Finding a formulation of general relativity closely resembling classical electrodynamics would allow to leverage the substantial recent progress made in electrodynamics in curved space, including balding of electromagnetic hair and ringdown \cite{Lyutikov:2011tk,Bransgrove:2021heo,Most:2024qgc}, reconnection and dissipation \cite{Parfrey:2018dnc,Comisso:2020ykg}, charged extremal black hole formation \cite{Kehle:2024vyt}, topological properties of the fields \cite{Wu:2024dao}, and in the case of neutron star binaries nonlinear interaction of the field prior to merger \cite{Palenzuela:2013hu,Most:2020ami,Most:2023unc,Kim:2024fuy}.

Re-incorporating some of the progress made in those areas might have the added potential to shed new light onto modifications of general relativity \cite{Cayuso:2023xbc}, as some of the challenges in modifying and incorporating new physics into Maxwell-like equations has progressed rapidly over the past years \cite{Denicol:2019iyh,Most:2021uck}.


Here we address this issue by providing a first analysis of spacetime collisions in terms of Maxwell-like electric and magnetic fields representing gravity itself.
Going beyond special or limit cases explored, e.g., in the context of the classical double-copy \cite{Monteiro:2014cda,Chawla:2023bsu,Caceres:2025eky}, we use a general formulation of Einstein equations applicable to the dynamical nature of a merging binary black hole system \cite{Olivares_2022}.

We do so by using a formulation of the Einstein equations originally due to Ref. \cite{Estabrook:1996wa}, with later modifications by Ref. \cite{Buchman:2003sq}. This set of equations has recently been demonstrated to be {\it exactly} equivalent to a nonlinear set of Maxwell equations \cite{Olivares_2022} (see also Ref. \cite{Peshkov:2022cbi}).This work uses $G=c=1$ units.\\

{\bf Basic picture.} In this work, we want to understand the dynamics of binary black hole collisions through analogies with relativistic electrodynamics. As such, we will exploit that general relativity can be interpreted as a gauge theory \cite{Baez:1995sj}, although crucially we will adopt a different formulation than the one commonly used \cite{Maartens:1997fg}.

The canonical starting point is the Einstein-Hilbert action,
\begin{align}\label{eqn:EH}
    S_{\rm EH} = \int {\rm d}x^4 \sqrt{-g} R\,,
\end{align}
where $R$ is the Ricci scalar, and $g$ is the determinant of the four-metric, $g_{\mu\nu}$.
The Einstein equations are then obtained by variation of the action with respect to the inverse metric, $g^{\mu\nu}$,
\begin{align}\label{eqn:Einstein}
    G_{\mu\nu} = 8 \pi T_{\mu\nu}\,,
\end{align}
where we have added a matter stress-energy tensor $T_{\mu\nu}$, and $G_{\mu\nu}$ is the Einstein tensor.\\
While this form of the Einstein equations is most commonly used, it obscures connections with its natural Yang-Mills character that enables interpretations in terms of generalized electric and magnetic fields \cite{Baez:1995sj}.
In turn, it is therefore more convenient to adopt a different formulation of the Einstein equations in local (tetrad) coordinates \cite{RevModPhys.48.393}. To this end, we define a local tetrad basis, $A^{\hat{\alpha}}{}_\mu$ such that
\begin{align}
    A^\mu{}_{\hat{\alpha}} A^\nu{}_{\hat{\beta}} g_{\mu\nu} = \eta_{\hat{\alpha}\hat{\beta}}\,,
\end{align}
where $\eta_{\hat{\alpha}\hat{\beta}}$ is the Minkowski metric. Here, $\hat{\alpha}$ is an index in the locally flat tetradic frame.
In order to ensure proper transformation between different local tetrad bases, the covariant derivatives need to be augmented by a spin connection \cite{RevModPhys.48.393}, $\omega^{\hat{\beta}}{}_{\hat{\alpha}\mu}$,
such that,
\begin{align}
    D_\mu V^\nu{}_{\hat{\alpha}} = \partial_\mu V^\nu{}_{\hat{\alpha}} + \Gamma^\nu{}_{\mu\kappa} V^\kappa{}_{\hat{\alpha}} + \omega^{\hat{\beta}}{}_{\hat{\alpha}\mu} V^\nu{}_{\hat{\beta}}\,,
\end{align}
is fully covariant.
One can then recast Eq. \eqref{eqn:EH} into Einstein-Palatini form \cite{Palatini:1919ffw},
\begin{align}\label{eqn:palatini}
    S_P = \int {\rm d}x^4 A^\mu{}_{\hat{\alpha}} A^\nu{}_{\hat{\beta}} \mathcal{F}^{\hat{\alpha}\hat{\beta}}{}_{\mu\nu}\,,
\end{align}
where $\mathcal{F}^{\hat{\alpha}\hat{\beta}}{}_{\mu\nu}$ is a Yang-Mills-like field strength tensor in the spin connection,
\begin{align}
    \mathcal{F}^{\hat{\alpha}\hat{\beta}}{}_{\mu\nu} = \partial_\mu \omega^{\hat{\alpha} \hat{\beta}}{}_{\nu} 
    - \partial_\nu \omega^{\hat{\alpha} \hat{\beta}}{}_{\mu}
    + \left[\omega,\omega\right]^{\hat{\alpha} \hat{\beta}}{}_{\mu\nu}\,,
\end{align}
where $\left[\omega, \omega\right]^{\hat{\alpha} \hat{\beta}}{}_{\mu\nu} = \omega^{\hat{\alpha} }_{\hat{\kappa}}{}_{\left[\mu\right|} \omega^{\hat{\kappa} \hat{\beta}}{}_{\left|\nu\right]}$ is a commutator. It is now possible to introduce electric and magnetic fields, $\mathcal{E}^{\hat{\alpha} \hat{\beta}}{}_{\mu} = n^\nu \mathcal{F}^{\hat{\alpha}\hat{\beta}}{}_{\mu\nu} $ and $\mathcal{B}^{\hat{\alpha} \hat{\beta}}{}_{\mu} = n^\nu\,^{\ast}\!\mathcal{F}^{\hat{\alpha}\hat{\beta}}{}_{\mu\nu}$, as effective components of this curvature tensor.
Here, $n_\mu$ is a timelike normal vector characterizing the time foliation of spacetime \cite{supplemental}.
A related starting point is geometrodynamics as a decomposition of the Weyl tensor \cite{Baez:1995sj}, and has been used to interpret back hole collision dynamics using vortices and tendices \cite{Owen_2011}. While the resulting equations themselves are Maxwell-like, with even a Poynting theorem for energy fluxes \cite{bel1959introduction}, the gravitational electric and magnetic fields obtained in this way are fundamentally different from classical electrodynamics, in the sense that for a single black hole with mass $M$, e.g., $\mathcal{E}^{\hat{r} \hat{r}}{}_{0} \sim -2 M /r^3$ is a tidal field \cite{Owen_2011}, and not a Coulomb field.

It would be desirable to introduce electric and magnetic fields more akin to classical electrodynamics, see e.g., Refs. \cite{Monteiro:2014cda} for a recent approach in the context of the classical double copy. To this end, we adopt here the DGREM formulation of Einstein's equations of Ref. \cite{Olivares_2022}, which we will summarize in the following. In order to introduce well-defined electric and magnetic fields, the field strength tensor should not  be expressed in terms of curvature directly, but in terms of the tetrad transformation, $A^{\hat{\alpha}}{}_\mu$ itself, which we promote to a vector potential. First, we introduce a field strength tensor directly related to the local tetrad frame,
\begin{align}
    F^{\hat{\alpha}}{}_{\mu\nu} = \partial_\mu A^{\hat{\alpha}}{}_\nu - \partial_\nu A^{\hat{\alpha}}{}_\mu\,.\label{eq:field_strength_tensor}
\end{align}
As we will see, $F^{\hat{\alpha}}{}_{\mu\nu}$ partially takes the role of a Maxwell field strength tensor, trivially giving rise to magnetic and electric fields, via 
\begin{align}
B^{\hat{\alpha}\mu} &= n_\lambda {}^{\ast}\!F^{\hat{\alpha}\mu\lambda}=\epsilon^{\mu\nu\kappa} \partial_\nu A^{\hat{\alpha}}{}_\kappa\,\label{eq:B_tensor},\\
E^{\hat{\alpha}\mu} &= n_\lambda F^{\hat{\alpha}\mu\lambda}\,\label{eq:E_tensor},
\end{align}
in the usual way. It is these electric and magnetic fields that will not only behave Maxwell-like, i.e., we will show that a Schwarzschild black hole has $E^{\hat{0}r}\sim M/r^2$, but will satisfy an almost exact set of Maxwell equations \cite{Olivares_2022}.

One of the striking features of this definition is that metric compatibility of the spin connection, $\omega^{\hat{\alpha} \hat{\beta}}{}_{\mu}$, implies a Christoffel-symbol-like form \cite{RevModPhys.48.393},
\begin{align}
    \omega_{\hat{\alpha}\hat{\beta}\hat{\gamma}}
= \frac{1}{2} \left( F_{\hat{\beta}\hat{\alpha}  \hat{\gamma}} + F_{\hat{\gamma}\hat{\alpha}  \hat{\beta}} - F_{\hat{\alpha}\hat{\beta}\hat{\gamma}}\right)\,.
\end{align}
which is not given in terms of $\partial_\mu A^{\hat{\alpha}}{}_\nu$ alone, but in terms of $F^{\hat{\alpha}}{}_{\mu\nu}$.

What is missing from the above description is how to connect the resulting expression to the Einstein equations \eqref{eqn:Einstein}, which in the language of electrodynamics requires us to produce an evolution equation for $E^{\hat{\alpha}\mu}$.
Following \cite{Olivares_2022}, we do so by using the Nester-Witten form \cite{Frauendiener:2006dx}, 
\begin{align}
{}^\ast\!\mathcal{U}^{\hat{\beta}\hat{\gamma}}{}_{\hat{\alpha}} = \omega^{[\hat{\beta}\hat{\gamma}]}{}_{\hat{\alpha}} + \delta^{\hat{\beta}}{}_{\hat{\alpha}} \omega^{[\hat{\gamma}\hat{\delta}]}{}_{\hat{\delta}} - \delta^{\hat{\gamma}}{}_{\hat{\alpha}} \omega^{[\hat{\beta}\hat{\delta}]}{}_{\hat{\delta}}\,,
\end{align}
which has the important property that,
\begin{align}\label{eqn:G_NW}
    G^\mu{}_{\hat{\alpha}} = \frac{1}{\sqrt{-g}} \partial_\nu \left(\sqrt{-g}\, {}^\ast\!\mathcal{U}^{\mu\nu}{}_{\hat{\alpha}} \right) - t^\mu{}_{\hat{\alpha}}\,,
\end{align}
where $t^\mu{}_{\hat{\alpha}}$ is a space-time current (see below). This is in analogy to the Landau-Lifshitz pseudo-tensor \cite{Clough:2021qlv}.
Using this definition, one can show that the Einstein-Palatini action \eqref{eqn:palatini} is equivalent up to boundary terms to \cite{Olivares_2022}
\begin{align}
    S_{\rm YM} = \frac{1}{4} \int {\rm d}x^4  {F}^{\hat{\alpha}\mu\nu} \,^\ast\!\mathcal{U}_{\mu\nu \hat{\alpha}}\,,
\end{align}
which closely resembles a canonical Yang-Mills action. In fact, the space-time current, $t^\mu_{\hat{\alpha}}$, in tetrad coordinates parallels the Maxwell stress-energy tensor \cite{Olivares_2022},
\begin{align}\label{eqn:stress_maxwell}
    t^{\hat{\beta}}{}_{\hat{\alpha}} = F^{\hat{\gamma}}{}_{\hat{\kappa}\hat{\alpha}}{}^\ast\!\mathcal{U}^{\hat{\kappa}\hat{\beta}}{}_{\hat{\gamma}} - \frac{1}{4} \delta^{\hat{\beta}}{}_{\hat{\alpha}} F^{\hat{\gamma}}{}_{\hat{\kappa}\hat{\lambda}}{}^\ast\!\mathcal{U}^{\hat{\kappa}\hat{\lambda}}{}_{ \hat{\gamma}}\,.
\end{align}
Using the Einstein equations \eqref{eqn:Einstein} together with Eq. \eqref{eqn:G_NW} results in two Maxwell equations,
\begin{align}\label{eqn:dgrem1}
     \partial_{\mu}\left(\sqrt{-g}\, {}^\ast\! F^{\mu\nu}{}_{\hat{\alpha}}\right) &=0\,,\\
     \partial_\mu\left(\sqrt{-g}\, {}^{\ast}\mathcal{U}^{\mu\nu}{}_{\hat{\alpha}} \right)&=\sqrt{-g}(t^\nu{}_{\hat{\alpha}}+ 8 \pi T^\nu{}_{\hat{\alpha}})\label{eqn:dgrem2}\,.
\end{align}
\begin{figure}[!t]
    \centering
    \includegraphics[width=\linewidth]{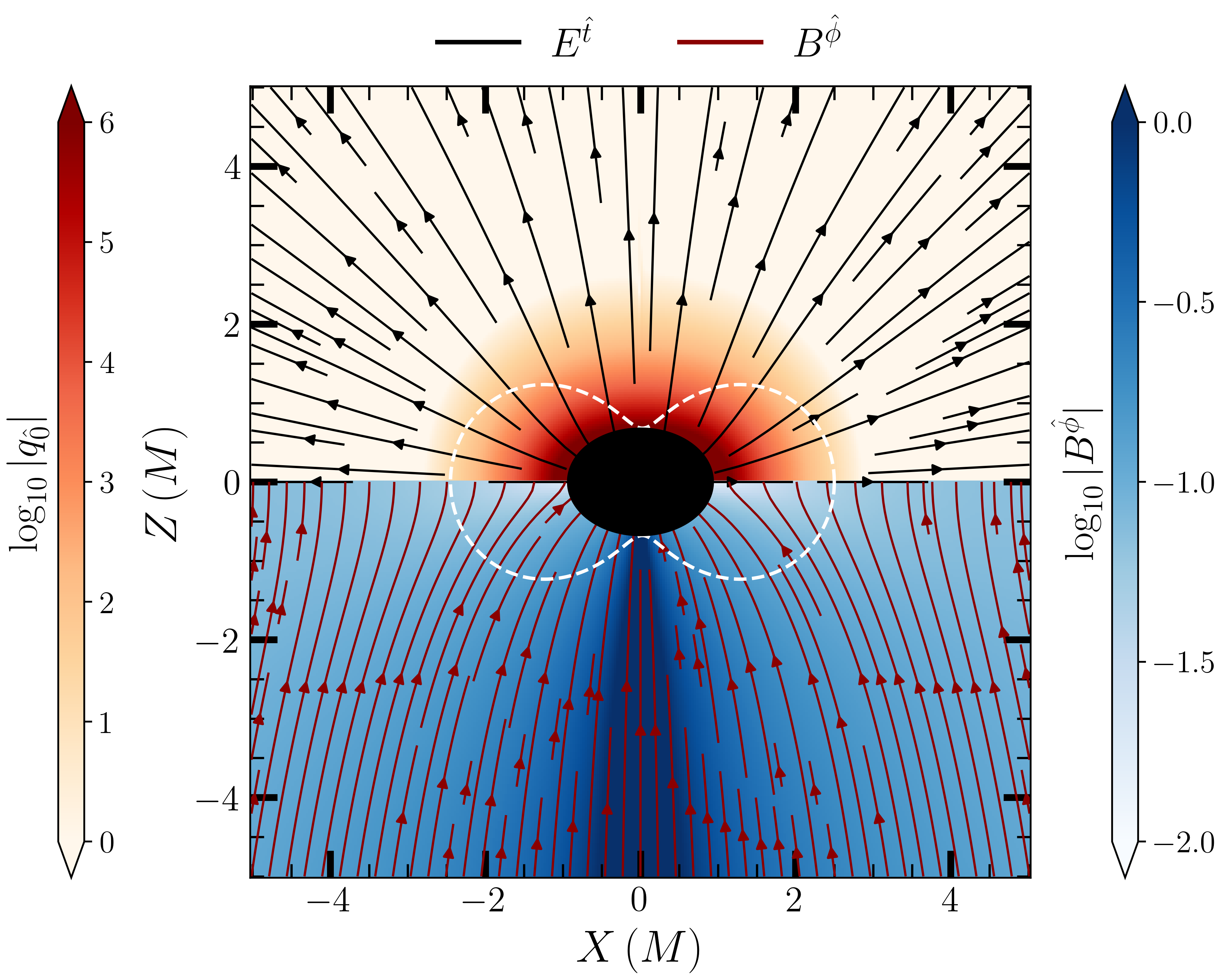}
    \caption{Rapidly spinning Kerr black hole in a dynamically evolved moving puncture gauge. Shown in color are the gravitational energy density $q_{\hat{0}}$, and toroidal magnetic field, $B^{\hat{\phi}\mu}$, in the meridional plane.
    Streamlines indicate the static component of the gravitational electric field, $E^{\hat{t}\mu}$, and magnetic fields. The black ellipse marks the black hole horizon, white dashed contours denote the ergosphere.}
    \label{fig:fig1_kerr}
\end{figure}
These equations perfectly resemble those of relativistic electrodynamics \cite{Baumgarte:2002vv}.
If we now interpret $\mathcal{U}^{\mu\nu}{}_{\hat{\alpha}}$ as a field strength tensor in its own right, we are led to introduce auxiliary electric and magnetic fields, 
\begin{align}
    H^{\hat{\alpha}\mu} \left[E,B\right] &= n_\lambda \mathcal{U}^{\mu\lambda\hat{\alpha}}\,,\\
D^{\hat{\alpha}\mu} \left[E,B\right]&= n_\lambda {}^{\ast}\!\mathcal{U}^{\mu\lambda\hat{\alpha}}\,,
\end{align}
which resemble electromagnetic fields in a medium, and are related to $B^{\hat{\alpha}\mu}$ and $E^{\hat{\alpha}\mu}$ via constitutive relations \cite{Olivares_2022},
\begin{align}
    H_{\hat{k}\hat{i}} &= - B_{\hat{i}\hat{k}} + \frac{1}{2} \delta_{\hat{k}\hat{i}} B^{\hat{l}}{}_{\hat{l}} - \epsilon_{\hat{k}\hat{i}\hat{l}} E_{\hat{0}}{}^{\hat{l}}\,,\\
    E^{\hat{j}}{}_{\hat{k}} &= - D_{\hat{k}}{}^{\hat{j}} - \frac{1}{2} \delta^{\hat{j}}{}_{\hat{k}} D^{\hat{l}}{}_{\hat{l}} + \epsilon^{\hat{j}}{}_{\hat{k}\hat{l}} H_{\hat{0}}{}^{\hat{l}}\,.
\end{align}
We can recast Eqs. \eqref{eqn:dgrem1}, \eqref{eqn:dgrem2} {\it exactly} in a more familiar vector calculus form \cite{Olivares_2022}, also following \cite{Komissarov:2004ms},
\begin{align}
\partial_t {\sqrt{\gamma} {\bf B}}^{\hat{\alpha}} + \nabla\times \left[ {\bf {E}}^{\hat{\alpha}}-{\boldsymbol \beta}\times{\bf B}^{\hat{\alpha}}\right] &=0\,\\ \label{eq:maxwell_equation_vector_form}
\partial_t {\sqrt{\gamma}{\bf D}}_{\hat{\alpha}} - \nabla\times \left[{\bf {H}}_{\hat{\alpha}}-{\boldsymbol \beta}\times{\bf D}_{\hat{\alpha}} \right] &= - \alpha \left[{\bf j}_{\hat{\alpha}}+ 8\pi{\bf J}_{\hat{\alpha}}\right] + \boldsymbol{\beta}q_{\hat{\alpha}}\,,\\
\nabla\cdot {\bf {B}}^{\hat{\alpha}} &=0\,,\\
\nabla\cdot {\bf {D}}_{\hat{\alpha}} &= q_{\hat{\alpha}} + 8\pi Q_{\hat{\alpha}}\,, \label{eqn:poisson}
\end{align}
where $\sqrt{\gamma} = \sqrt{-g}/\alpha$, and $\alpha$ and $\boldsymbol{\beta}$ are the gravitational lapse and shift, respectively \cite{supplemental}.
In doing so, we have also introduced spacetime ($q_{\hat{\alpha}}/{\bf j}_{\hat{\alpha}}$), and matter ($Q_{\hat{\alpha}}/{\bf J}_{\hat{\alpha}}$) charges/currents which follow directly from the stress-energy tensor \eqref{eqn:stress_maxwell},
\begin{align}\label{eqn:qhat}
    q_{\hat 0} &= -\frac{\sqrt{\gamma} }{2}\left[\boldsymbol{ E\cdot D + B \cdot H}\right]\,,\\
    q_{\hat{i}} &= - \sqrt{\gamma} \left(\boldsymbol{B\times D}\right)_{\hat{i}}\,, \
\end{align}
and, the spacetime current is given by
\begin{align}
   {\bf j}_{\hat{0}} &= -\sqrt{\gamma} {\bf E}\times{\bf H}\,, \label{eqn:Poynting}\\
   {\bf j}_{\hat{i}} &= \sqrt{\gamma} \left({\bf E :D}_{\hat{i}} + {\bf B :H}_{\hat{i}} + q_{\hat{0}}{\bf I}_{\hat{i}}  \right)\,,
\end{align}
and $Q_{\hat{\alpha}}$ and ${\bf J}_{\hat{\alpha}}$ is the external energy density and matter current, which we do not consider in this work.
Here ${\bf A}\cdot{\bf B} = A_{\hat{\alpha}\hat{j}} B^{\hat{\alpha}\hat{j}}$, and ${\bf A} : {\bf B}_{\hat{i}} = A_{\hat{\alpha}}{}^{\hat{k}}B^{\hat{\alpha}}{}_{\hat{i}} $.
From the above equations it is already apparent that many of the concepts from classical electrodynamics carry over, including electrostatics of point charges via a Poisson-like equation \eqref{eqn:poisson}, or Poynting's theorem for energy fluxes \eqref{eqn:Poynting}. We point out that a numerical implementation of the constraint sector would require techniques akin to constraints transport \cite{Evans:1988qd} or divergence cleaning \cite{dedner2002hyperbolic} approaches, as discussed in Ref. \cite{Olivares_2022}.\\

{\bf Black hole electrostatics.}
In order to gain some intuition into the nature of the electric and magnetic fields describing spacetime itself, we first explore analytic examples for isolated non-rotating and rotating black holes.

\begin{figure}[!t]
    \centering
    \includegraphics[width=0.95\linewidth]{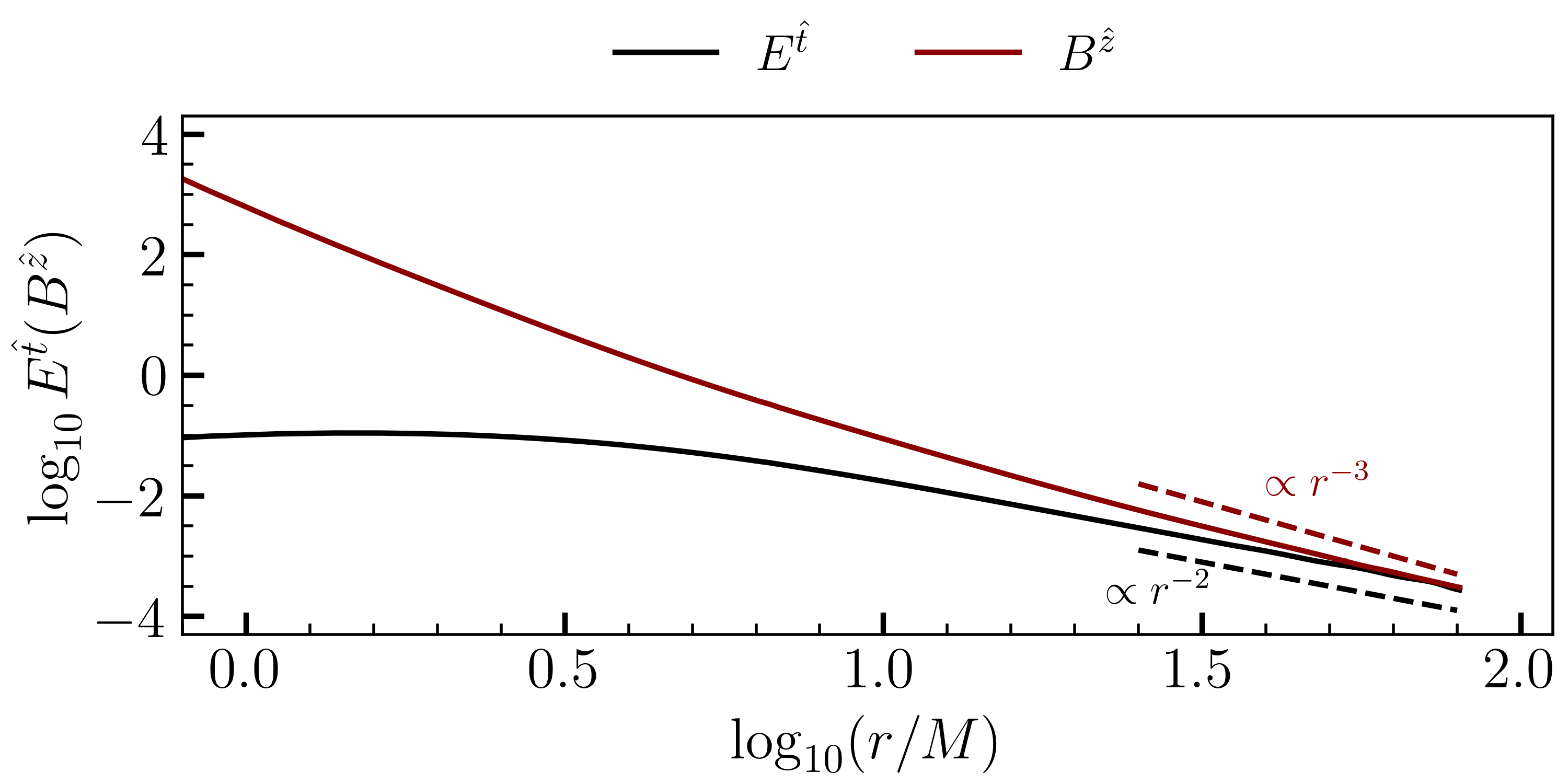}
    \caption{Far-field behavior of the gravitational electric and magnetic fields of a single Kerr black hole.}
    \label{fig:fig2_Kerr_scaling}
\end{figure}

A non-rotating Schwarzschild black hole in Schwarzschild coordinates is described by 
\begin{align}
    {\rm d}s^2 = -\left(1-\frac{r_s}{R}\right)\text{d}t^2 + \left(1-\frac{r_s}{R}\right)^{-1}\text{d}R^2+R^2\text{d}\Omega^2,
\end{align} 
where $r_s = 2M$ is the Schwarzschild radius, and $R = \sqrt{x^2 + y^2 + z^2}$.

\begin{figure*}
    \centering
    \includegraphics[width=0.9\linewidth]{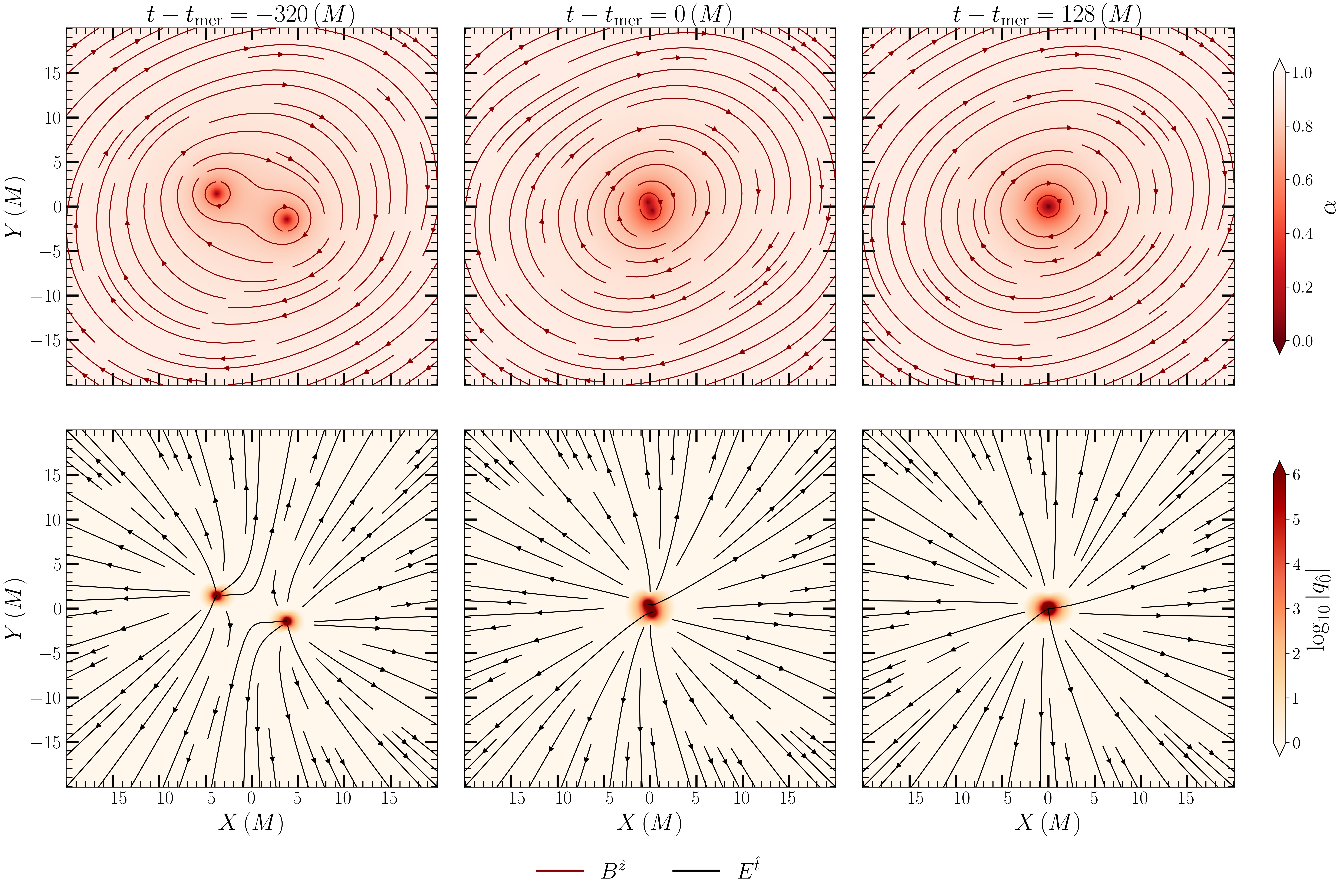}
    \caption{Gravitational electric and magnetic field dynamics of a binary black hole merger in the inspiral, merger and ring-down phases on the orbital plane. The first row shows the gravitational lapse $\alpha$ and magnetic field lines $B^{\hat{y}\mu}$. The second row shows the energy density $q_{\hat{0}}$ and electric field lines $E^{\hat{t}\mu}$. Times, $t$, are stated relative to the time of merger, $t_{\rm mer}$.}
    \label{fig:fig3_BBH_zoomin}
\end{figure*}

The diagonal form of the metric naturally allows for a simple choice for tetrad,
$A^{\hat{t}}{}_{\nu} = \left(1-r_s/R\right)^{1/2}\delta^{0}{}_\nu$,
as well as, $A^{\hat{r}}{}_{\nu} =\left(1-r_s/R\right)^{-1/2}\delta^{1}{}_\nu,\,A^{\hat{\theta}}{}_{\nu}=R\delta^{2}{}_\nu,\,A^{\hat{\phi}}{}_\nu=R\sin{\theta}\delta^{3}{}_\nu$.
It is straightforward to show that the only non-vanishing component of the electric field \eqref{eq:E_tensor} is
\begin{align}\label{eqn:schwarzschild_efield}
    E^{\hat{t} R} = \frac{M}{R^2}\,.
\end{align}
This implies that -- in full analogy to classical electrostatics --  the black hole has a Coulomb electric field with charge $M$.
We can further compute the force exerted by this field, by computing the effective Lorentz force, ${\bf F}$, which can be computed as \cite{Olivares_2022},
\begin{align}
    {F}_{\hat{i}} = {\bf E}_{\hat{i}}\cdot {\bf j}_{\hat{0}} 
    + E^{\hat{0}}{}_{\hat{i}} q_{\hat{0}} + \left({\bf {j}} \times {\bf {B}}\right)_{\hat{i}}\,,
\end{align}
which fully mimics the classical Lorentz force, i.e., $\mathbf{F} = \rho \mathbf{E} - \mathbf{J}\times \mathbf{B}$, where $\rho$ is the charge density, and $\mathbf{J}$ the electric current. In particular we find, that the electrostatic contribution, $F_{\hat{R}} \simeq E^{\hat{t}}_{R} q_{\hat{0}} =q_{\hat{0}} M/R^2 $, is exactly Coulomb's law.

However, as we can see this is not the only contribution to the Lorentz force as the magnetic field in the chosen tetrad coordinates is nonvanishing \cite{supplemental}. Indeed, we find that $B\simeq M/R^3$, resembling a magnetic dipole field in radial dependence, but not geometry, which is in part toroidal.

We can further extend this analysis to a rotating black hole, see also Ref. \cite{Maluf:2023rwe} for a related analysis in Fermi-Walker tetrads. As such, we consider a rotating Kerr black hole with mass $M$ and dimensionless spin $\chi = 0.9$ \cite{Kerr:1963ud}.  In order to prepare ourselves for the numerical analysis of the binary black hole spacetime to follow, we now adopt a numerical tetrad \cite{supplemental}. This requires to use a form of the Kerr metric suitable for numerical integration, and we adopt quasi-isotropic coordinates \cite{Liu:2009al}. We then evolve the spacetime using the Z4c formulation of Einstein's equations in moving puncture gauge \cite{Bernuzzi:2009ex}, as implemented in the \texttt{Frankfurt/IllinoisGRMHD} code \cite{Most:2019kfe}, which operates on top of the \texttt{Einstein Toolkit} \cite{Loffler:2011ay}. We then use this numerical solution of the Einstein system to reconstruct the gravitational electric and magnetic fields in post-processing.\\
\begin{figure*}
    \centering
    \includegraphics[width=\linewidth]{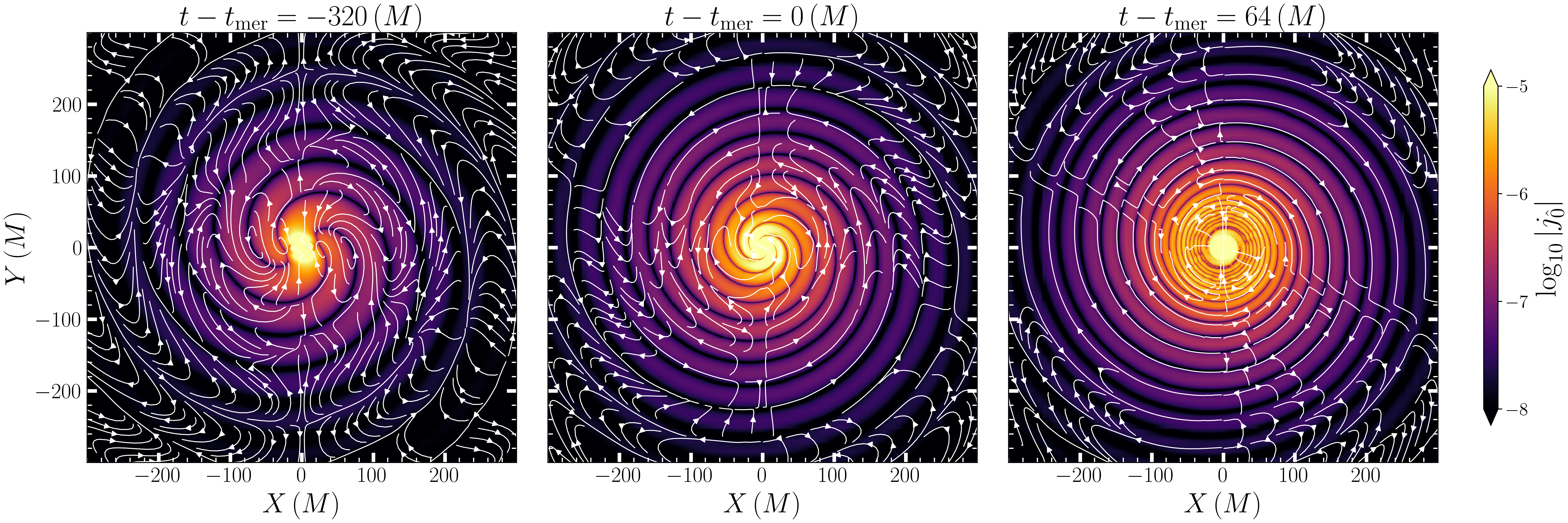}
    \caption{Gravitational Poynting flux magnitude $j_{\hat{0}}$ in the orbital plane at the same times as Fig.~\ref{fig:fig3_BBH_zoomin}. White lines denote the gravitational electric field, ${\bf E}^{\hat{r}}$, indicating gravitational wave radiation.}
    \label{fig:fig4_BBH_current}
\end{figure*}
The resulting stationary solution is shown in Fig. \ref{fig:fig1_kerr}.
We explicitly show the spacetime energy/charge density, see Eq.~\eqref{eqn:qhat}.
We can see that the spacetime charge density is almost exclusively distributed inside and near the black hole horizon. The electric field is radial, and we demonstrate that it reproduces the expected fall-off at large distances in Fig. \ref{fig:fig2_Kerr_scaling}. The magnetic field has a Wald like geometry, akin to a black hole embedded in a uniform magnetic field \cite{Wald:1974np}. However, the magnetic field is still penetrating the black hole horizon for this rapidly spinning black hole, unlike the prediction for a Meissner effect for a vacuum magnetic field \cite{PhysRevD.12.3037}. The field lines seem most strongly affected inside the ergosphere, although (and as expected for a stationary solution) we do not see the formation of any dissipative region as would be present in a magnetospheric Wald solution \cite{Komissarov_2004}. The fall-off is consistent with the far-field limit obtained from the Kerr solution \cite{supplemental}, as shown in Fig.~\ref{fig:fig2_Kerr_scaling}.
{\bf Binary black hole electrodynamics. }
We now focus on electrodynamics aspects of a binary black hole collision. We consider an equal mass non-spinning system with an initial separation $a=11\,M$. The initial data is generated in the Bowen-York formalism \cite{Bowen:1980yu} using puncture initial data for the black holes \cite{Brandt:1997tf} on a post-Newtonian orbital trajectory \cite{Healy:2017zqj}. Numerically, this initial condition is computed using the \texttt{TwoPunctures} code \cite{Ansorg:2004ds}, and is evolved using the same numerical infrastructure as in the previous Section.
We begin by focusing on the inspiral and merger dynamics, see e.g., Ref. \cite{Lehner:2014asa} for an overview of the different stages of a binary black hole merger.
Initially, the two black holes orbit each other (Fig. \ref{fig:fig3_BBH_zoomin}). We can see that the gravitational electric field of the binary configuration strongly resembles that of two equal point charges in classical electrostatics. The gravitational electric fields emanating from the black holes repel each other along the dividing plane, highlighting the intrinsic mirror symmetry of the problem.
Additionally, the static component of the magnetic field (Fig. \ref{fig:fig3_BBH_zoomin}, top panel) is predominantly toroidal.
In fact, just like for a current going through two wires, the toroidal fields of both black holes merge into a common, bar-deformed magnetic field, which is maintained throughout the inspiral. At merger, the two black holes collide, leading to a transient in the gravitational electric and magnetic fields (center panel) in order to propagate out the large bar deformation ($m=2$ mode) from the inspiral. The final black hole then is a perturbed Kerr black hole, as described above.

On the electrodynamics side, the orbiting black holes source the emission
of gravitational waves. 
Rather than considering the gravitational wave signal itself
\citep{Bishop:2016lgv}, we focus here on understanding and interpreting the
gravitational wave flux in terms of electrodynamic analogies. 

In Fig.~\ref{fig:fig4_BBH_current}, we now define the energy-momentum current of the gravitational electromagnetic fields $j^{k}{}_{\hat{\mu}}$, which appears on the right-hand-side of Ampere's law, Eq.~\eqref{eq:maxwell_equation_vector_form}.
This expression is completely analogous to the Poynting flux density commonly used in electrodynamics \cite{Baumgarte:2002vv}.

Gravitational wave emission is sourced by a time-varying mass quadrupole
\cite{Peters:1964zz}. We can see this as an imprint on the gravitational
Poynting flux as a primary and secondary spiral radiation component
starting from the orbital light cylinder.
Overall, this structure resembles a quadrupolar version of a striped wind
component of a pulsar (e.g., \cite{Petri:2012cs}, which also exists for
orbiting neutron star binaries \cite{Carrasco:2020sxg,Palenzuela:2013hu,Most:2020ami,Most:2023unc,Kim:2024fuy}). We can also
recognize the quadrupolar wave pattern in the gravitational electric field,
${\bf E}^{\hat{r}}$. In fact, we can directly spot that subsequent wave
cycles are clearly separated only at large distances ($r\gtrsim 200M$), and
are interacting at smaller distances. This visually illustrates the
difference between near-zone and wave zone behavior of the gravitational
waves, only in the latter of which gravitational waves can be thought of as
propagating in empty space \cite{Bishop:2016lgv}.\\ At merger the frequency
of the emission increases as can be seen in the center panel of Fig.
\ref{fig:fig4_BBH_current}, after which the dynamical component of the
electric field rapidly becomes stationary, coincident with the
high-frequency ring-down (right panel).

\begin{figure}
   \centering
   \includegraphics[width=0.95\linewidth]{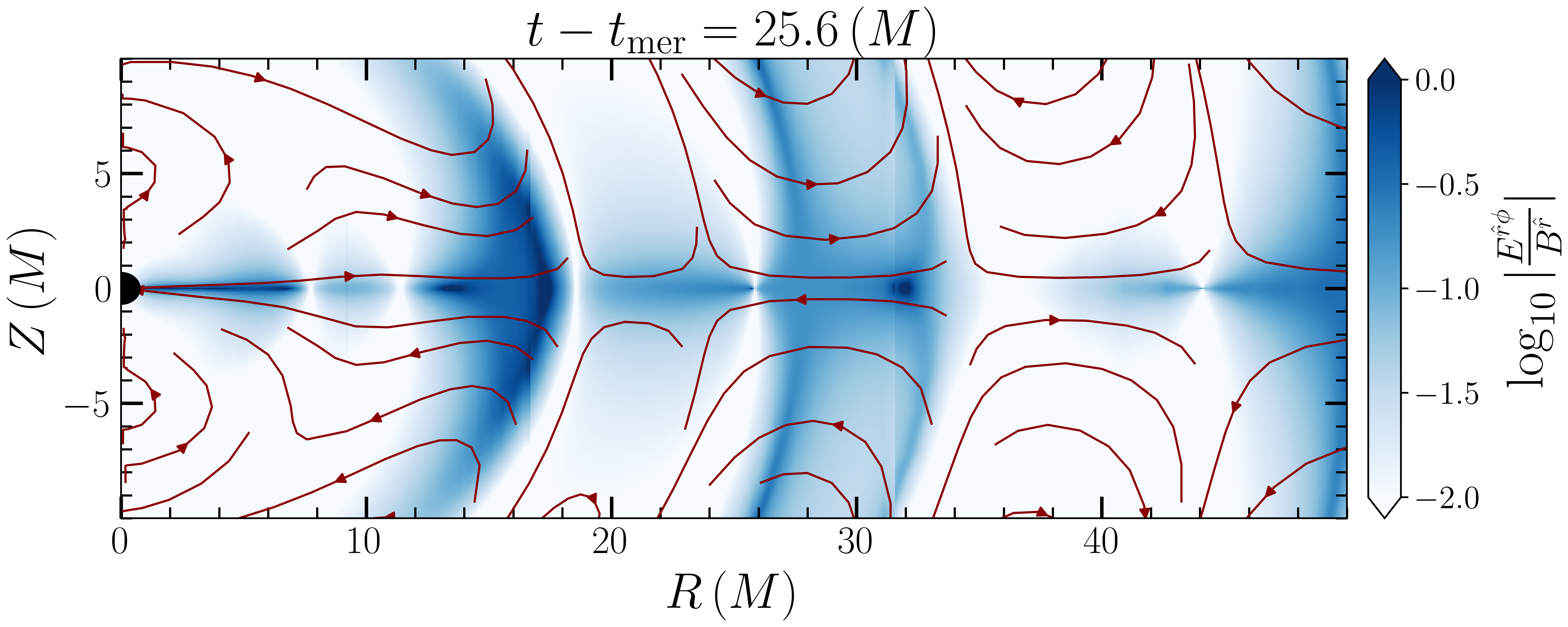}
   \caption{Gravitational wave ring-down signature from the merged black hole (black disk). The ratio of the gravitational electric field component $E^{\hat{r}\phi}$ and magnetic field strength $B^{\hat{r}}$ is plotted on the meridional plane.}
   \label{fig:fig5_BBH_ring-down}
\end{figure}

The ring-down of a black hole is a well studied phenomenon \cite{Teukolsky:1973ha,Berti:2009kk}, and has undergone recent interest in particular related to black hole quasi-normal mode spectroscopy \cite{Giesler:2019uxc}.
These quasi-normal modes can also be imprinted in a regular electromagnetic field \cite{Teukolsky:1973ha,Baumgarte:2002vu}, which will exhibit the same quasi-normal decay as the spacetime. As a special case, it has been shown that black holes can source quasi-normal fast modes when embedded into a plasma \cite{Most:2024qgc}. Here we show that a similar morphology also holds for the gravitational electrodynamic field. In analogy to fast modes, we defined an effective velocity $v\simeq E^{\hat{r}\phi}/\left|B^{\hat{r}}\right|$ of the ring-down waves. In Fig. \ref{fig:fig5_BBH_ring-down} we demonstrate that these well resemble similar quasi-normal imprints in their classical electromagnetic analogue \cite{Most:2024qgc}.

{\bf Conclusions.}
In this work, we have provided a new visualization and interpretation of the dynamical spacetime of a binary black hole collision.
By using a Palatini-like formulation of the Einstein equations in a local tetrad frame \cite{Olivares_2022}, we have analytically and numerically computed the gravitational electric and magnetic fields of a single and a binary black hole spacetime. 
In particular we analytically and numerically confirm a Coulomb law for electrostatic fields of a stationary black hole spacetimes. In addition, we find these spacetimes to have strong toroidal magnetic fields. During the dynamical collision of a binary black hole spacetime, we identify gravitational wave emission in the near and far zone, as well as ring-down imprints in the gravitational electric and magnetic fields, akin to their electrodynamic counterparts \cite{Baumgarte:2002vu,Most:2024qgc}.

While exploratory, our initial study highlights the strength of using electrodynamics formulations to interpret general relativity spacetimes. Natural extensions of this work would be a detailed analysis of the ring-down including a quantification of nonlinearities \cite{Cheung:2022rbm,Mitman:2022qdl}, as well as a connection of the tetrad to the BMS symmetry group \cite{Freidel:2021fxf}.
Finally, there has been tremendous progress in modeling dissipative effects in relativistic plasmas using Israel-Stewart-like equations \cite{Denicol:2019iyh,Most:2021uck}. In how far these approaches can be carried over or combined with existing formulations of modified theories of gravity would be another interesting avenue \cite{Cayuso:2023xbc}.

\begin{acknowledgments}
ERM is grateful for discussions with Cynthia Keeler, Mark Scheel and Anatoly Spitkovsky.
SB acknowledges support through Caltech's SURF program. JW acknowledges partial support through the David and Barbara Groce graduate fellowship. ERM acknowledges partial support by the National Science Foundation under grants No. PHY-2309210 and AST-2307394.
ERM acknowledges the use of Delta at the National Center for Supercomputing Applications (NCSA) through allocation PHY210074 from the Advanced Cyberinfrastructure Coordination Ecosystem: Services \& Support (ACCESS) program, which is supported by National Science Foundation grants \#2138259, \#2138286, \#2138307, \#2137603, and \#2138296. Additional simulations were performed on the NSF Frontera supercomputer under grant AST21006. ERM also acknowledges support through DOE NERSC supercomputer Perlmutter under grant m4575, which uses resources of the National Energy Research Scientific Computing Center, a DOE Office of Science User Facility supported by the Office of Science of the U.S. Department of Energy under Contract No. DE-AC02-05CH11231 using NERSC award NP-ERCAP0028480.
\end{acknowledgments}

\nocite{*}
\bibliography{reference}

\end{document}